# CLAUDE BOUCHU, INTENDANT DE BOURGOGNE AU 17ÈME SIÈCLE, A-T-IL INVENTÉ LE MOT « STATISTIQUE » ?

Dominique PEPIN[*]

RÉSUMÉ

L'objet de cet article est d'examiner la thèse selon laquelle le mot « statistique » aurait été utilisé pour la première fois au 17ème siècle, dans un mémoire écrit par Claude Bouchu, intendant de Bourgogne. Une analyse historique et bibliographique est menée pour juger la crédibilité de cette thèse. L'inspection physique du mémoire permet ensuite d'apporter une réponse définitive.

ABSTRACT

The objective of this paper is to examine the assertion that the word "statistics" would have been used for the first time in the 17th century, in a report written by Claude Bouchu, administrator of Bourgogne. A historical and bibliographical analysis is carried out to judge the credibility of this thesis. The physical inspection of the report then makes it possible to bring a final answer.

## 1. Introduction

Alors que l'activité statistique de recueil des données est ancienne, et même plusieurs fois millénaire[1], le mot *statistique* est apparu beaucoup plus récemment. Il est né sous la plume d'un auteur allemand inconnu, qui publie en 1672 sous le pseudonyme d'Helenus Politanus l'ouvrage : *Microscopium Statisticum : quo status imperii romano-germanici cum primis extraordinarius, ad vivum repraesentur*, écrit en latin germanique (Hecht, 1997, et Dupaquier et Dupaquier, 1985). Le mot *statisticum* est alors utilisé par Helenus Politanus sous la forme d'un adjectif. Il fut par la suite substantivé, et connut quelques variations[2] (*statisticae, statistica, statistik,* …).

---

[*] Institut des Risques Industriels Assurantiels et Financiers, Centre Du Guesclin – Place Chanzy, 79000 Niort
E-mail : Dominique.Pepin@univ-poitiers.fr

[1] Les premiers recensements et enregistrements d'événements démographiques, constituant ce que Dupaquier et Dupaquier (1985, chapitre 1) appellent « les balbutiements de la statistique », remontent à la Haute Antiquité. Il existe par exemple pour la Mésopotamie des statistiques militaires vieilles de presque 5000 ans.
[2] On peut consulter Dupaquier et Dupaquier (1985, p. 20-22) pour connaître le détail de ces évolutions.

Pour l'Ecole allemande, qui a permis la diffusion du mot *statistique*, et dont Gottfried Achenwall (1719-1772), professeur de droit international et de science politique à Goettingue, fut un éminent représentant, le mot *statistique* n'avait pas le sens que nous lui prêtons aujourd'hui. Desrosières (1993, p. 219) résume ainsi la définition originelle de ce terme :

> « Le mot "statistique" est né en Allemagne au XVIII$^è$ siècle pour désigner une "science de l'Etat" (staatwissenschaft) ».

Ce mot désignait la science de la constitution de l'Etat, et procédait par des descriptions de type littéraire, décrivant les forces physiques, morales et politiques des divers Etats composant l'Europe. La statistique allemande constituait une science humaine mêlant géographie et science politique, bien différente de l'approche statistique moderne adoptée par l'école anglaise de *l'arithmétique politique*. L'arithmétique politique anglaise, inventée par William Petty[3] (1623-1687), véritable ancêtre de la statistique moderne, hérita finalement du mot *statistique*, passé d'Allemagne en Angleterre par l'intermédiaire de Zimmermann, professeur à Brunswick. Ce nouveau mot (*statistics*) remplaça rapidement l'expression d'*arithmétique politique*.

Cette lecture de l'histoire du mot *statistique*, retenue par la grande majorité des historiens, est aujourd'hui remise en cause par quelques auteurs (notamment Meyer, 1981), qui contestent l'attribution à Helenus Politanus de l'invention dudit mot. En effet, la redécouverte d'un mémoire écrit au 17$^{ème}$ siècle par l'intendant de Bourgogne Claude Bouchu laisse penser que ce dernier aurait utilisé le mot *statistique* quelques années avant la parution du *Microscopium Statisticum* de Helenus Politanus. Ce mot serait ainsi peut-être d'origine française. Les recherches sur la première apparition écrite de ce terme sont ainsi relancées, et elles conduisent à de multiples interrogations : qui était Claude Bouchu ? Quelles sont les preuves ou faisceaux de preuves indiquant qu'il aurait inventé le mot *statistique* ? A-t-il réellement inventé ce mot ?

A ces diverses questions, nous apportons des éléments de réponse, présentés de la façon suivante : en premier lieu, nous décrivons rapidement le rôle de l'intendant, et dressons un portrait succinct de l'homme, Claude Bouchu. Nous présentons ensuite le mémoire et les indications tendant à prouver que l'intendant aurait inventé le mot *statistique*. Puis nous nous interrogeons quant à la pertinence de ces preuves ; nous sommes alors conduits à nous interroger quant à la validité du titre présumé du mémoire de Bouchu, puisque c'est dans cet intitulé qu'apparaîtrait le mot *statistique*. Finalement, Claude Bouchu a-t-il vraiment inventé ce mot ? Nous tranchons la question dans la dernière partie.

---

[3] William Petty est pour cette raison considéré par certains auteurs contemporains comme le premier statisticien de l'histoire. Il nous faut aussi mentionner Graunt (1620-1674), dont les travaux sur les bulletins de décès ont inspiré Petty, et Davenant (1656-1714), qui a poursuivi l'œuvre de Petty.

# 2. Claude Bouchu, intendant de Bourgogne de 1656 à 1683

Le nom de Claude Bouchu n'est sans doute pas familier au lecteur, sauf s'il est un passionné de l'histoire de Bourgogne, ou s'il habite Dijon (une rue porte son nom). En effet, l'histoire de Bouchu est attachée à la région de Bourgogne, dont c'est un personnage historique notoire, et plus encore à la ville de Dijon. Qu'il soit nous soit permis ici de présenter rapidement l'intendant Bouchu, d'après les éléments à notre disposition.

Après lecture de son bulletin de baptême et de son acte de décès[4], nous apprenons que Claude Bouchu, chevalier, conseiller d'Etat ordinaire du Roy en tous ses conseils, Comte de Pont de Veyle, marquis des Essarts et autres lieux, fut intendant de Bourgogne pendant 27 ans, de 1656 à 1683. Né à Dijon, en Côte-d'Or, le quatre mai 1628, il y mourut le huit juin 1683, à l'âge de 55 ans.

Pour décrire ce singulier personnage et sa place dans l'histoire de la Bourgogne, renvoyons à Arbassier (1919, p. 1-2), qui ne se cache pas d'en être un fervent admirateur et lui a consacré une thèse de doctorat :

> « Ce qui réhausse la personnalité de Bouchu, et fait de lui un homme digne de l'histoire, c'est que cet homme est véritablement une pierre dans un édifice.
>
> L'individu ne laisse une trace qu'autant il s'est insinué dans l'évolution de la pensée ou de la civilisation. Les génies n'ont été que des continuateurs et des devanciers, de puissants collaborateurs à la destinée humaine toujours en marche. Bouchu n'a pas été un génie, mais il a eu le mérite de coopérer de toute son énergie à un grand moment de notre histoire nationale. Avoir contribué à l'absolutisme royal dans un coin de la France, la Bourgogne, voilà son titre à l'attention. »

Pour comprendre la place de Bouchu dans la société bourguignonne de l'époque[5], nous pouvons présenter l'intendant comme un représentant du pouvoir royal, à volonté centralisatrice, dont l'objectif est de briser toute velléité d'indépendance provinciale. « Fort outil d'assimilation monarchique que l'intendance », écrivent à juste titre Calmette et Drouot (1928, p. 283). Arbassier (1919, p. 16) décrit le rôle de l'intendant en les termes suivants :

> « Aux temps où nous sommes, les intendants sont devenus, en effet, les factotums de ce pouvoir. Celui-ci les envoie à demeure dans les provinces pour qu'ils y intriguent à son profit, au préjudice des libertés locales, au bénéfice du pays. »

Pour Desrosières (1993, p. 38), « le rôle et le comportement des intendants annoncent ceux des préfets des $XIX^e$ et $XX^e$ siècles ». L'intendant Bouchu représente en Bourgogne le bras juridique, fiscal, politique et financier du roi, comme le soulignent Barbero et Brunet (1978a, p. 10) :

---

[4] La lecture de ces documents nous a été nécessaire pour certifier sa date de naissance, à cause d'une équivoque due à Arbassier (1919), spécialiste de Bouchu, qui mentionne à tort 1627 pour son année de naissance, et qui se trompe de surcroît d'une dizaine de jours sur la date de sa mort.
[5] Nous renvoyons à Calmette et Drouot (1928, chapitre XIV) le lecteur désireux de connaître plus en détails l'histoire de la Bourgogne sous le règne de Louis XIV.

« Intendant de Justice, il surveille tous les officiers de son ressort, peut présider tous les tribunaux et instituer des commissions extraordinaires ; Intendant de Police, il maintient l'ordre, s'occupe des problèmes de subsistance et des ponts et chaussées, surveille les municipalités ; Intendant des Finances, il veille à la répartition et à la levée de la taille par les Elus et les Trésoriers de France en pays d'élection et soumet aux Etats provinciaux les exigences royales en pays d'Etat. »

On comprend aisément l'intérêt du pouvoir royal d'avoir à la tête de chaque région un fidèle intendant qui lui soit entièrement dévoué. Or Bouchu est par essence cet intendant fidèle et infaillible du roi et de son ministre Colbert (Calmette et Drouot, 1928, p. 283-284) :

« L'intendant de Colbert, l'intendant-type, bon chef de file, c'est Bouchu, un fils de premier président (…), élevé même par un père « principion ». Bourgeois obstiné et travailleur, homme à tout faire du souverain, alternativement souple et brutal, il administre avec passion durant vingt-huit ans (…), reflète visiblement à Dijon les volontés, les desseins du ministre dont il est à bien des égards comme la réplique réduite et provinciale, et qu'en 1683 il accompagne fidèlement dans la mort »

Arbassier (1919, p. 23) partage ce point de vue et insiste sur l'opiniâtreté de l'intendant :

« Du coup apparaît en plein jour la vraie personnalité de Bouchu : au fond, il n'est jamais que l'homme à tout faire de la monarchie, l'agent qui emploie son zèle à mettre la province à l'entière disposition du Roy, tout comme s'il ne s'agissait que d'une simple terre de Boussolles. Ce but, il le poursuit avec la dernière fermeté. La force l'accompagne comme une inséparable servante. Quand un obstacle résiste à son habileté, il le brise (…). On dépasserait la vérité si l'on allait jusqu'à appliquer à Bouchu l'épithète venimeuse et bien saint-simonienne de garde-chiourne. »

Mais que l'on ne s'y trompe pas, si Bouchu est entièrement dévoué à la cause royale, cela n'est pas non plus sans arrière-pensée, sans aucune recherche d'intérêt personnel. Par sa fidélité et son dévouement, il espère gagner les faveurs du roi, et avec l'espoir de s'attirer bien des sympathies, il manie avec perfection l'art de la flatterie. Arbassier (1919, p. 10) commente ainsi la conduite de ce « galant homme » :

« Bouchu fut-il habile ou influent ? Il sut certainement plaire et séduire. La loyauté n'exclut pas l'affabilité entre égaux et les démonstrations respectueuses qu'on doit aux grands. »

Il poursuit sa description, nous le présentant comme un véritable thuriféraire du pouvoir (Arbassier, 1919, p. 17) :

« C'est pourquoi nous voyons Bouchu sans cesse appliqué à fortifier son crédit, à plaire aux grands et tout d'abord aux ministres. Qualité professionnelle : il ne leur arrive aucune faveur qu'il n'applaudisse ».

Nous pouvons résumer le portrait de Claude Bouchu en le présentant comme un « agent zélé du pouvoir royal » (Barbero et Brunet, 1978a, p. 11), comme un infatigable travailleur qui a mis toute son énergie à exécuter les volontés du roi et de son ministre, Colbert.

# 3. Le mémoire de Bouchu et la première trace écrite du mot *statistique*

Après avoir peint succinctement le portrait de Claude Bouchu, intéressons-nous au processus qui l'aurait conduit à inventer le mot *statistique*. Pour cela, décrivons cet événement particulier qu'a été la rédaction du mémoire de l'intendant de Bourgogne.

Pour comprendre cette invention, replaçons-nous dans le contexte administratif français du 17$^{\text{ème}}$ siècle. L'information dont dispose le roi sur l'état des provinces (richesses, nombre d'habitants,…) est très sommaire, et ce manque est perçu comme un frein à l'action et au développement (Meyer, 1981, p. 221-222) :

> « Finalement, nous retrouvons toujours le même problème : la rationalisation du système de gouvernement repose sur la connaissance, la prise de conscience de la portée réelle des moyens d'action de l'Etat du pays.
>
> D'où la manie des enquêtes. La condition du grand dessein est la connaissance. Colbert est l'homme de l'information, qu'il systématise à grande échelle, à un degré absolument inconnu de ses prédécesseurs. »

Plus que tout autre, Colbert perçoit le besoin d'une description détaillée des provinces de la France. Il est alors tout naturel de faire appel aux intendants de ces provinces (Desrosières, 1993, p. 38-39) :

> « (…) Ces intendants sont chargés de faire parvenir au roi des descriptions de leurs provinces selon des modalités de plus en plus codifiées. Remontant à la tradition médiévale de « miroir du prince », destiné à instruire celui-ci et à lui présenter le reflet de sa grandeur, c'est-à-dire de son royaume, extension métaphorique de son propre corps, ce système d'enquêtes va peu à peu se dédoubler, d'une part, en un tableau descriptif et général réservé au roi et, d'autre part, en un ensemble de connaissances particulières, quantifiées et périodiques, destinées aux administrateurs. »

En 1665, Colbert fait entreprendre par ses intendants une vaste enquête dans les provinces de France[6]. D'après Barbero et Brunet (1978a, p.9), « Claude Bouchu fut commis en lettres patentes du 22 septembre 1665 pour réaliser cette enquête dans l'intendance de Bourgogne ». Gille (1980), se référant à Esmonin (1956), et Deveze (1960), font état d'une circulaire qui aurait été envoyée par Colbert dès mars 1664 dans les diverses provinces, pour demander aux intendants de faire cette enquête. Cette circulaire est confirmée par l'ordonnance royale du 7 août 1664 et l'arrêt du conseil d'Etat du 7 août 1665 (Garnier, 1880).

Dans la circulaire de mars 1664, Colbert demandait aux intendants de dresser un tableau complet du clergé, de la noblesse, d'examiner la conduite des membres du parlement, de fournir des renseignements sur les officiers de finance, sur le montant et la nature des impôts royaux. Il exprimait aussi son désir de supprimer les dettes des communautés, et demandait une étude de la situation du commerce, des manufactures, de l'agriculture, ainsi que du prix des principales marchandises.

---

[6] Cette grande enquête a été précédée par une enquête de moindre ampleur, confiée par Colbert en avril 1663 à son frère, Charles Colbert de Croissy, et concernant les provinces d'Alsace, de Lorraine et des Trois Evêchés (Gille, 1980).

De l'enquête de l'intendant de Bourgogne naît un mémoire, intitulé : *Déclaration des biens, charges, dettes et statistique des communautés de la généralité de Dijon*[7]. Le mot *statistique* serait ainsi né de la plume de Bouchu ; ce serait en tout cas la première trace écrite du mot.

Décrivons à présent ce mémoire, qui se distingue nettement des autres mémoires rédigés par les intendants. Appuyons-nous sur la description qu'en donne Deveze (1960, p.86) :

> « Pour la Bourgogne, notre mémoire se présente sous une forme encore beaucoup plus éloignée des textes de synthèse que constituent les autres mémoires. C'est une suite de questionnaires, paroisse après paroisse, sans qu'aucune synthèse vienne couronner le tout. Classées par baillages, les paroisses et communautés de Bourgogne sont au nombre de 1950. C'est donc une masse énorme de 7.800 pages, à raison de 4 par communauté, formant neuf volumes imposants, aux armes de France et Navarre. Cette matière considérable n'a pas été mise en œuvre, semble-t-il, à moins qu'un ouvrage abrégé composé par l'intendant Bouchu n'ait été perdu. »

Le zèle de Bouchu l'a conduit à mener une enquête aux proportions démesurées (Barbero et Brunet, 1978a, p. 9-10) :

> « La forme sous laquelle il a consigné les réponses obtenues s'éloigne assez nettement de la synthèse demandée par Colbert, et sous bien des aspects elle dépasse par son ampleur le projet original ».

Bien entendu, l'intendant de Bourgogne n'a pu mener à lui tout seul cette enquête. Il délégua ses fonctions à des officiers (les subdélégués) chargés chacun d'une circonscription réduite. Bouchu fit appel aux subdélégués pour faire l'enquête et rédiger le mémoire. A ce sujet, Deveze (1960, p. 97-98) écrit :

> « Rien ne nous permet de connaître par le mémoire lui-même les noms et fonctions de ceux qui ont participé à sa rédaction : tout ce qu'on peut dire, c'est que les 7.800 pages de questionnaires sont couvertes par une dizaine d'écritures différentes. »

Sur le rôle des subdélégués, et la façon dont Bouchu a recouru à leur aide, nous pouvons consulter Moreau[8] (1948, p. 165-166) :

> « Un subdélégué est donc « en général celui qui a été commis pour agir sous les ordres d'un officier supérieur » (Guyot [1784]). Le mot de commission, qui s'oppose habituellement à celui d'office, désigne une charge temporaire et révocable qui cesse à la volonté du commettant. Ainsi la subdélégation, à ses débuts, présente toujours un caractère exceptionnel et de circonstance. En outre le subdélégué, de quelque autorité qu'il relève, n'a pas de pouvoir qui lui soit propre ; il agit au nom d'une personne ou d'une compagnie qui lui confèrent un pouvoir limité en vue d'accomplir

---

[7] Notons l'emploi du mot statistique sans « s ». Le titre complet du mémoire du Bouchu, ou du moins celui archivé par Garnier (1880), archiviste en chef des archives civiles départementales de Côte-d'Or, est le suivant : *Déclaration des biens, charges, dettes et statistique des communautés de la généralité de Dijon, fournies par ordre de Bouchu, intendant de Bourgogne, et sous forme de questionnaire, d'après les procès-verbaux dressés dans chaque communauté par les subdélégués, de 1666 à 1669, en conformité de l'ordonnance royale du 7 août 1664 et de l'arrêt du conseil d'Etat du 7 août 1665.*
[8] Moreau (1948) cite lui-même Guyot (1784).

une mission déterminée et ce service ne lui vaut que très rarement des indemnités ; en tout cas, jamais de gage. Son pouvoir est essentiellement précaire et temporaire. »

Le mémoire de Bouchu, sans doute à cause de son imposant volume, n'a jamais été édité[9]. Il n'en existe que quelques manuscrits, que Barbero et Brunet (1978a, p. 16-18) ont recensés au nombre de quatre :

« Tout d'abord, pour l'ensemble des baillages de la Généralité de Bourgogne, il existe un exemplaire à la Bibliothèque Nationale et un autre aux Archives Départementales de la Côte-d'Or présenté en 9 volumes in-folio faisant 7800 pages ; texte des questions imprimé et réponses manuscrites. Ensuite, pour les bailliages de Bresse, Bugey et Gex uniquement, il existe un exemplaire à la Bibliothèque Municipale de Bourg (2 tomes, 434 et 488 feuillets) et un autre aux Archives Départementales de l'Ain (666 et 720 feuillets). L'exemplaire de la bibliothèque est semblable dans sa présentation aux précédents, tandis que le dernier cité est entièrement manuscrit (questions et réponses).

Même si l'écriture de l'exemplaire des Archives Départementales de l'Ain est manifestement plus récente, il est certain que les quatre exemplaires ont une origine commune (…).

Certains indices peuvent donc laisser penser que l'exemplaire conservé à la Bibliothèque Municipale de Bourg aurait servi de base aux autres copies, plus ou moins tardives, mais aucune preuve ne permet de l'affirmer définitivement. »

En plus de ces exemplaires manuscrits, il faut aussi mentionner la publication par Barbero et Brunet (1978a, 1978b) des résultats de l'enquête de Bouchu pour les seuls baillages de Bugey, du Pays de Gex et de Bresse.

Mais quand la rédaction du mémoire de Bouchu a-t-elle précisément pris fin ? Le titre complet du mémoire, tel qu'il est archivé par Garnier (1880), laisse entendre que les procès-verbaux de l'enquête auraient été dressés entre 1666 et 1669. La rédaction du mémoire n'aurait alors pu être achevée avant 1669. Pour Barbero et Brunet (1978a), l'enquête aurait eu lieu entre 1666 et 1670, et la rédaction n'aurait donc pu être terminée avant 1670. La correspondance de Bouchu avec Colbert laisse toutefois penser que la rédaction du mémoire aurait été plus rapide. En prenant appui sur cette correspondance, Deveze (1960, p.83) écrit :

« Il a été rédigé pour l'essentiel en 1665. En effet, on se réfère généralement pour chaque paroisse aux impositions de l'année 1665, et pour un certain nombre d'entre elles seulement à celles de 1666. L'intendant de Dijon, Bouchu, fait allusion dans sa correspondance avec Colbert à ce gros travail en cours, notamment dans une lettre du 29 novembre 1665 ; le 16 avril 1666, il y travaillait encore, il a terminé le 2 juin. »

D'après cette correspondance, on peut conclure que Bouchu aurait achevé (avec l'aide de ses subdélégués) en juin 1666 la rédaction du mémoire, dont le titre comporte le mot *statistique*. Mais quelle signification Bouchu accordait-il à ce mot ? Vu le titre et la forme du mémoire, il semblerait que le mot *statistique* ait été utilisé dans son acception moderne, c'est-à-dire comme un ensemble de données[10]. Bouchu aurait été de façon surprenante très en avance sur son temps[11], et la première trace écrite du mot en français précéderait la première trace écrite latine du mot, apparue quelques années plus tard (1672) sous la plume de Helenus Politanus.

---

[9] Après consultation et manipulation des neuf volumes du mémoire de Bouchu, nous pouvons attester que chacun d'entre eux pèse une dizaine de kilos. Chaque volume fait une soixantaine de centimètres dans sa longueur, une quarantaine dans sa largeur, et une quinzaine de centimètres d'épaisseur.

Evidemment, qui ne voudrait pas en savoir plus ? Comment serait venu à Bouchu l'idée d'utiliser ce nouveau mot ? Quelqu'un lui aurait-il suggéré son utilisation ? Après consultation des neuf volumes du mémoire, nous pouvons garantir que le mot *statistique* n'apparaît que dans le titre, et à aucun autre endroit. Le mémoire ne nous donne donc aucune indication sur le sens et l'origine de ce mot. Ces éléments invitent selon nous à la prudence. Il paraît en effet surprenant que Bouchu ait utilisé un nouveau mot dans le titre de son mémoire, sans donner par la suite aucune explication sur ce terme, et cela en lui donnant une acception semblable à celle qu'on lui prête aujourd'hui.

## 4. L'intitulé du mémoire de Bouchu

A présent que nous avons décrit le mémoire rédigé à l'intention de Colbert et du roi, interrogeons-nous sur la pertinence des éléments laissant penser que l'intendant de Bourgogne aurait inventé le mot *statistique*.

A la fin de la section précédente, nous avons exprimé quelques réserves quant à la thèse qui présente Bouchu comme l'inventeur de ce mot. Mais comment peut-on douter de cette proposition, qui semble avérée ? Comme nous l'avons signalé précédemment, le mot *statistique* n'apparaît que dans le titre du mémoire. Si le titre originel du mémoire est celui présumé, alors Bouchu est bien l'inventeur de ce mot. Si au contraire l'intitulé d'origine n'est pas celui annoncé, et ne comporte pas le mot *statistique*, alors l'intendant n'a pas inventé ce terme.

La solution à ce problème est simple : il suffit juste d'aller consulter les exemplaires manuscrits du mémoire de Bouchu et de regarder quel titre ils portent, ce que nous avons fait. Mais avant de nous exprimer sur ce que nous avons trouvé, voyons d'abord comment les auteurs qui ont mentionné l'enquête de Bouchu ont cité son mémoire. Cela nous éclairera sur la réponse finale, et sur le sens à lui donner.

Selon nous, il ne fait pas de doute que l'historien qui a permis la propagation de l'idée que Bouchu aurait inventé le mot *statistique* est Meyer (1981). Aussi penchons-nous d'abord sur ce qu'il a écrit au sujet du mémoire de Bouchu (Meyer, 1981, p. 225) :

---

[10] Aujourd'hui, lorsqu'on parle des données et non de la discipline qui analyse ces données, on parle de *statistiques* (au pluriel), alors que Bouchu utilise le singulier, terme utilisé de nos jours pour désigner la discipline qui se propose d'analyser ces données.

[11] Rappelons que par la suite, le mot *statistique* a été utilisé pendant plusieurs décennies par l'Ecole allemande dans un sens différent de son acception moderne, avant de trouver celui-ci grâce à l'école anglaise de l'arithmétique politique.

> « De 1666 à 1669 l'intendant de Bourgogne Claude Bouchu dressa une « déclaration des biens, charges, dettes et *statistiques*[12] des communautés de la généralité de Bourgogne[13] ». Le vocabulaire est intéressant : tous les dictionnaires attribuent le mot statistique au dictionnaire de Trévoux[14] (XVIIIè siècle), du latin *statisticum* (qui a trait à l'Etat), germanisé par Schmeitzel en 1749, à moins que ce ne soit Achanwall[15]. Il n'en est rien : le mot appartient au langage administratif français colbertien ».

Remarquons combien Meyer (1981) est affirmatif. Pour lui, il n'y a pas de doute : le mot *statistique* est bien d'origine française. Il n'emploie même pas à cet égard le conditionnel. On comprend alors l'influence qu'il a pu avoir sur certains de ses lecteurs. Ainsi, Droesbeke et Tassi (1997) se réfèrent à Meyer (1981) et mentionnent à sa suite le mémoire de Bouchu où le mot statistique apparaîtrait pour la première fois. Le lecteur y apprend que « le mot « statistique » appartient au langage administratif français colbertien » (Droesbeke et Tassi, 1997, p. 39).

Meyer (1981) et Droesbeke et Tassi (1997) ne sont pas pas les premiers à citer le mémoire de Bouchu. Bien avant eux, Roupnel (1922), dans son étude sur les populations du pays dijonnais, évoque le mémoire de Bouchu, mais sans jamais citer le titre présumé de l'ouvrage. Il parle seulement à de nombreuses reprises de « l'enquête ordonnée par Bouchu ». Gille (1980), dans son ouvrage *les sources statistiques de l'histoire de France : des enquêtes du XVIIè siècle à 1870*, parle à peine d'un vague mémoire sur la Bourgogne, alors qu'il mentionne la grande enquête lancée par Colbert. Tout au plus écrit-il à ce sujet (Gille, 1980, p.25) :

> « Tout récemment, M. Deveze, a signalé une description sociale, administrative, économique et financière de la Bourgogne, rédigée par paroisse. Il y a donc là un travail extrêmement précis et détaillé, qui aurait servi à la rédaction d'un mémoire sur la Bourgogne qui ne vit jamais le jour ».

Cela montre combien le mémoire de Bouchu est resté longtemps méconnu. Pas plus Esmonin (1956) que les autres ne cite cet ouvrage, alors qu'il s'intéresse pourtant aux mémoires des intendants. Deveze (1960, p. 77-79) nous confirme de sa plume que le mémoire de Bouchu n'est pas ou très peu connu des historiens, même des historiens spécialistes de Colbert ou de la Bourgogne :

> « Or, on n'a pas tiré parti, jusqu'à présent, d'une manière suffisante, de ce mémoire original. Il semble inconnu aux historiens du village sous l'ancien régime, comme Albert Babeau, qui s'occupe surtout, il est vrai, de la Champagne. Il paraît ignoré des auteurs d'une histoire admnistrative de la France, comme Chéruel, ou des auteurs d'une histoire des intendants de province, comme Hanotaux et Charles Godard. Il n'est pas connu davantage des historiens de Colbert, dont les plus célèbres restent Clément et Lavisse. Les historiens du droit français, comme Brissaud, Chénon et Olivier-Martin, l'ont négligé, mais on ne peut leur en tenir rigueur, car les historiens de la Bourgogne eux-mêmes ont généralement oublié d'en faire mention (ainsi Noël Garnier, dans son introduction aux archives civiles de la Bourgogne). Kleinclausz, puis Drouot et Calmette, dans leur *Histoire de la Bourgogne*, connaissent certainement notre mémoire, mais comme leurs ouvrages sont des synthèses destinées au grand public, ils ne le citent pas directement et l'utilisent à peine. M$^e$ Charles Arbassier, avocat à Dijon, qui avait consacré en 1921 à « l'intendant Bouchu et à son action financière » une intéressante thèse de droit, ne s'occupe que de la correspondance de l'intendant de 1667 à 1671, et fait seulement une brève allusion au mémoire sur les paroisses dans une courte note de bas de la page 12 de son livre, où il est question de la misère des communautés bourguignonnes à l'époque de Colbert. Alexandre Thomas lui-même, auteur d'une thèse de doctorat ès lettres sur la Bourgogne au temps de Louis XIV (*Une province sous Louis XIV*, 1844), n'a pas consulté un seul ouvrage de la Bibliothèque nationale, et n'a, semble-t-il, retenu de l'exemplaire du mémoire qui se trouve à Dijon que ce qui concerne les villes de Dijon, Beaune, Noyers, Arnay-le-Duc et Chalon. Aussi parle-t-il à peine des campagnes.

---

[12] En italique et au pluriel dans le texte de Meyer (1981, p. 225).
[13] On voit que « Dijon », apparaissant dans le titre archivé par Garnier (1880), est devenu ici « Bourgogne ».
[14] Le dictionnaire de Trévoux est un des premiers dictionnaires de la langue française. Il a été édité pour la première fois en 1704 à Trévoux. Il existe huit éditions de ce dictionnaire, la dernière datant de 1771. Après consultation de la dernière édition, il nous est apparu que ni le mot *statistique*, ni sa version latine *statisticum*, n'y sont référencés.
[15] Meyer (1981, p. 225) écrit bien « Achanwall » et non « Achenwall ».

Seul, l'illustre historien de la Bourgogne, Gaston Roupnel, a su profiter dans son ouvrage, La ville et la campagne au XVII$^è$ siècle, des données du mémoire de 1665 sur les collectivités locales de Bourgogne. Il considère ce mémoire dans sa bibliographie critique comme un document de premier ordre. Mais il n'en a effectivement étudié qu'un tome sur dix, celui qui est consacré au baillage de Dijon. »

Comme Deveze (1960) l'a signalé (voir la citation supra), Arbassier (1919) lui-même, dont la thèse de doctorat est pourtant consacrée à Claude Bouchu, mentionne à peine le mémoire de l'intendant. Il lui prête le titre « Déclaration des biens, charges, dettes et statistiques des communautés, fournies par ordre de Bouchu, intendant de Bourgogne », ou du moins il se réfère aux archives départementales de Dijon, série C, Intendance, C. 2882 à C. 2891, où le mémoire de Bouchu était classé sous ce titre (Arbassier, 1919, p. 186). Quant à Deveze (1960), il parle du mémoire de Bouchu ou des questionnaires de l'intendant Bouchu sur la Bourgogne, mais il ne lui prête jamais aucun titre faisant référence au mot *statistique*. Enfin, plus récemment, Barbero et Brunet (1978a et 1978b), lesquels ont édité dans leur *Déclaration des biens des communautés, 1665-1670*, une partie du mémoire de Bouchu, ne lui donnent non plus aucun titre.

Alors finalement, le mémoire de Bouchu comportait-il un titre faisant mention du mot *statistique* ? Notre réponse est catégorique et négative. Il n'apparaît aucun titre sur la couverture des manuscrits, ni dans les pages intérieures. Dans sa correspondance avec Colbert ou avec d'autres personnes[16], il n'y a aucune trace d'un titre quelconque donné par Bouchu à son mémoire. Mais d'où vient alors ce titre faisant référence au mot *statistique*, mentionné par tant d'auteurs ?

---

[16] On peut renvoyer le lecteur à Garnier (1895). Dans son ouvrage sur Dijon et la Bourgogne, il emploie intensivement la correspondance de Bouchu. La correspondance de Bouchu de 1667 à 1671 est conservée à Troyes, dans le Fonds Bouhier.

A l'origine de cette confusion, il y a le travail de l'archiviste Garnier (1880), responsable à Dijon des archives antérieures à 1790, qui procède à un classement de tous les ouvrages et manuscrits disponibles aux archives départementales de Côte-d'Or, et construit un inventaire-sommaire. Mais afin que tous les ouvrages et manuscrits apparaissent correctement dans cet inventaire-sommaire, encore faut-il qu'ils aient tous un titre permettant de les identifier et de les retrouver. Alors, aux ouvrages et manuscrits qui n'en possèdent point, Garnier leur donne un titre, apparaissant dans l'inventaire-sommaire, censé présenter la substance de l'ouvrage. Le titre présumé du mémoire de Bouchu est donc né de la plume de Garnier. C'est un attribut postérieur à des fins de classification.

Ce travail d'archivage devient la source d'une importante confusion, quand Arbassier (1919) cite le mémoire de Bouchu dans une note de bas de page. Il est très vraisemblable qu'Arbassier savait que le titre prêté par Garnier (1880) au mémoire de Bouchu n'était pas le titre original ; mais comme il se réfère à l'exemplaire possédé par les archives départementales de Côte-d'Or, il donne la référence telle qu'elle apparaît dans l'inventaire-sommaire construit par Garnier (1880).

Par la suite, beaucoup d'auteurs ne feront plus la distinction, simplement parce qu'ils n'auront pas l'occasion de consulter eux-mêmes le mémoire de Bouchu. Sans doute la confusion est-elle née avec la note de bas de page d'Arbassier (1919). Ceux qui ont lu sa thèse de doctorat mais qui n'ont pas consulté le mémoire de Bouchu n'ont pu comprendre que le titre cité n'était pas le titre original. Ensuite, par le biais des citations en chaîne, la confusion s'est installée, jusqu'à gagner Meyer (1981), qui prête au mémoire de Bouchu le titre inventé par Garnier.

L'intendant de Bourgogne n'était donc pas en avance sur son temps. Il n'a pas inventé le mot *statistique*. Il paraissait étrange qu'il ait pu utiliser ce mot dès 1666, au sens moderne du terme. En fait, le titre date de 1880, et cela explique une telle utilisation du mot *statistique*.

## 5. Conclusion

Nos connaissances quant à la première trace écrite du mot *statistique* ont été bousculées par quelques auteurs, se référant au mémoire de l'intendant de Bourgogne Claude Bouchu, et plus particulièrement à son titre, faisant mention du mot en question. On a alors pu croire que le mot *statistique* était né dans l'administration française colbertienne du XVII[ème] siècle. Qu'en est-il en vérité ? Claude Bouchu a-t-il vraiment inventé le mot *statistique* ?

Nous nous sommes efforcés dans cet article de répondre à cette question. Nous avons retrouvé les ouvrages et articles faisant référence au mémoire de Bouchu, et nous avons consulté les exemplaires manuscrits du mémoire.

Nous affirmons qu'il n'y a à ce jour aucune trace pouvant laisser croire que Claude Bouchu, ou quelque autre personnage de l'administration française colbertienne, ait pu inventer le mot *statistique*. Le mémoire de Bouchu ne comporte en fait aucun titre, et en conséquence pas de titre faisant référence au mot *statistique*. Il n'y a pas non plus dans le mémoire d'allusion à ce mot, pas plus qu'il n'y en a dans la correspondance de Bouchu.

Les éléments laissant croire que Bouchu aurait inventé ce mot sont erronées. C'est en fait Garnier (1880), responsable des archives antérieures à 1790 du département de Côte-d'Or, qui donne un titre au mémoire de Bouchu, afin de le classer dans son inventaire-sommaire, titre comportant alors le mot *statistique*. De cette intitulation postérieure à l'écriture du mémoire de Bouchu, est née une confusion, prêtant à l'intendant de Bourgogne la paternité d'une invention qui n'était pas la sienne. Nous espérons par ce travail avoir mis fin à cette méprise.

# RÉFÉRENCES